\documentclass[twocolumn]{IEEEtran} 

\usepackage{amsmath,amssymb,amsfonts}
\usepackage[final]{graphicx}
\usepackage{psfrag}
\usepackage[tight]{subfigure}
\usepackage[numbers,sort&compress]{natbib}
\usepackage{multicol}
\usepackage{mathrsfs} 
\usepackage{epstopdf}
\usepackage{multirow}
\usepackage{amsthm,mathtools}
\usepackage{float}
\usepackage{booktabs}  
\usepackage{multirow} 
\usepackage{relsize}
\usepackage{bm}
\usepackage{hyperref}
\usepackage{tabularx}

\usepackage[nolist,nohyperlinks]{acronym}
\usepackage{svg}
\usepackage{xparse}
\NewDocumentCommand{\evalat}{sO{\big}mm}{%
  \IfBooleanTF{#1}
   {\mleft. #3 \mright|_{#4}}
   {#3#2|_{#4}}%
}
\newcommand{\norm}[1]{\left\lVert#1\right\rVert}

\begin{acronym}
\acro{PLS}{physical layer security}
\acro{OP}{outage probability}
\acro{FRIS}{\emph{Fluid RIS}}
\acro{SNR}{signal-to-noise ratio}
\acrodefplural{SNR}[SNRs]{signal-to-noise ratios}
\acro{AWGN}{additive white Gaussian noise}
\acro{CSI}{channel state information}
\acro{PDF}{probability density function}
\acrodefplural{PDF}[PDFs]{probability density functions}
\acro{CDF}{cumulative distribution function}
\acro{PDF}{probability density function}
\acrodefplural{CDF}[CDFs]{cumulative distribution functions}
\acro{LRS}{large reflecting surface}
\acro{BS}{base station}
\acro{RF}{radio frequency}
\acro{RV}{random variable}
\acro{FN}{folded normal}
\acro{TDD}{time-division duplexing}
\acro{LOS}{line-of-sight}
\acro{DL}{downlink}
\acro{MISO}{multiple-input single-output}
\acro{MRT}{maximal ratio transmission}
\acro{MRC}{maximal ratio combining}
\acro{MC}{Monte Carlo}
\acro{RIS}{reconfigurable intelligent surface}
\acro{RV}{random variable}
\acro{CDF}{cumulative density function}
\acro{FAS}{fluid antenna system}
\acro{MIMO}{multipe-input multiple-output}
\acro{BS}{base station}
\acro{CSI}{channel state information}
\end{acronym}

\newtheorem{proposition}{Proposition}

\newcommand{\diag}{\mathop{\mathrm{diag}}}

\usepackage[numbers,sort&compress]{natbib}

\makeatletter
\def\blfootnote{\xdef\@thefnmark{}\@footnotetext}
\makeatother

\usepackage{float}
\usepackage{stfloats}
\usepackage{textcomp}
\usepackage{algpseudocode}
\usepackage{algorithm}

\newcommand{\comment}[1]{{\color{black} #1}}

\begin{document}
\title{\LARGE{Understanding the Role of Phase and Position Design in Fluid Reconfigurable Intelligent Surfaces}}

\author{J.~D.~Vega-S\'anchez, \textit{Senior Member, IEEE}, V. H. Garzón Pacheco, N. V. Orozco Garz\'on, \textit{Senior Member, IEEE},  H. R. Carvajal Mora, \textit{Senior Member, IEEE}, and F.~J.~L\'opez-Mart\'inez, \textit{Senior Member, IEEE} }

\maketitle

\blfootnote{\noindent Manuscript received MONTH xx, YEAR; revised XXX. The review of this paper was coordinated by XXXX. The work of F.J. L{\'o}pez-Mart{\'i}nez was supported by grant PID2023-149975OB-I00 (COSTUME) funded by MICIU/AEI/10.13039/501100011033 and FEDER/UE. \\
(\textit{Corresponding author: Jos\'e~David~Vega-S\'anchez})}

\blfootnote{\noindent 
J.~D.~Vega-S\'anchez, and H.~R.~Carvajal~Mora are with	Colegio de Ciencias e Ingenier\'ias  ``El Polit\'ecnico", Universidad San Francisco de Quito (USFQ), Diego de Robles S/N, Quito (Ecuador) 170157.
}

\blfootnote{\noindent
V. H. Garzón Pacheco, and N.~V.~Orozco~Garzón are with the Faculty of Engineering and Applied Sciences (FICA), Telecommunications Engineering, Universidad de Las Am\'ericas
(UDLA), Quito 170124, Ecuador. 
}

\blfootnote{\noindent
F.J. L{\'o}pez-Mart{\'i}nez is with Dept. Signal Theory, Networking and Communications, Research Centre for Information and Communication Technologies (CITIC-UGR), University of Granada, 18071, Granada, (Spain).
}

\vspace{-12.5mm}
\begin{abstract}
Fluid Reconfigurable Intelligent Surfaces (FRISs) are gaining momentum as an improved alternative over classical RIS. However, it remains unclear whether their performance gains can be entirely attributed to spatial flexibility, or instead to differences in  equivalent aperture or phase design. In this work, we shed light onto this problem by benchmarking FRIS vs. RIS performances in two practical scenarios: conventional RIS (same number of active elements and same overall aperture) and compact RIS (same number of active elements, and smaller aperture with sub-$\lambda$ inter-element spacing). Statistical analysis demonstrates that: (\textit{i}) spatial position optimization in FRIS provides noticeable gains over conventional RIS in the absence of phase-shift design; (\textit{ii}) such benefits vanish when FRIS and conventional RIS employ optimal beamforming (BF) and phase shift (PS) design, making position optimization irrelevant; (\textit{iii}) FRIS consistently outperforms compact RIS with optimized BF and PS design, owing to spatial correlation and smaller aperture.
\end{abstract}


\vspace{-2.5mm}
\section{Introduction}

\textcolor{black}{In the last years, }\ac{RIS} technologies have \textcolor{black}{carved out their place as a} solution to enhance wireless communication performance through cost-effective and energy-efficient control of the propagation environment. \textcolor{black}{Also recently, \ac{FAS} have been conceived as an alternative mechanism to enable antenna reconfigurability, by exploiting spatial oversampling to opportunistically mimic the effects of changing the shape and position of antenna elements \cite{Ref2}.} \textcolor{black}{While the benefits of the integration of \ac{RIS} and \ac{FAS} have been explored in the literature \cite{FASRIS},} the notion of \ac{FRIS} has been \textcolor{black}{newly} introduced ~\cite{Abdelhamid}. \textcolor{black}{FRIS architecture leverages} the physical relocation of reflecting elements over a predefined surface area, thus adding spatial configuration as an additional optimization dimension \textcolor{black}{over fixed-position RIS}. \textcolor{black}{The extension of the FRIS paradigm to simultaneously transmit and reflect signals further broadens the range of application scenarios \cite{Rostami}. } 
\textcolor{black}{The potential of FRIS in different use cases is being examined in different under various configurations and propagation conditions \cite{Xiao,Farshad}. However, while state-of-the-art studies predict notable improvements of FRIS implementations, the key benefits of position and phase optimization are not clearly delineated due to incomplete benchmarking.} 

\textcolor{black}{Motivated by the above, our aim in this paper is to provide a comprehensive performance analysis of FRIS-based systems. Specifically, we focus} on the fully optimized FRIS configuration where spatial position optimization, beamforming (BF), and phase shift (PS) are jointly designed. We benchmark this setup against a conventional RIS with identical numbers of \textcolor{black}{fixed-position} elements and optimized BF+PS, ensuring a fair comparison. \textcolor{black}{To assess the performance of the schemes under consideration}, we employ an Evolutionary Particle Swarm Optimization (E-PSO) algorithm for spatial position selection, a distributed optimization approach for BF and PS design, and an unsupervised learning-based method to approximate the optimized end-to-end channel. \textcolor{black}{Our results offer key insights into the conditions under which the spatial flexibility of FRIS yields substantial performance gains, and when these benefits become marginal.}
Our results reveal key insights into when the spatial flexibility of FRIS provides significant gains and when such advantages diminish.

\vspace*{2mm}
\noindent\textit{Notation and terminology:} Uppercase and lowercase bold letters denote matrices and vectors, respectively;$f_{(\cdot)}(\cdot)$ is a probability density function (PDF); $F_{(\cdot)}(\cdot)$ is a cumulative density function (CDF); $\mathcal{C}\mathcal{N}(\cdot ,\cdot )$ is the circularly symmetric complex Gaussian distribution; $J_0(\cdot)$ is the zeroth-order Bessel function of the
first kind; $\left ( \cdot \right )^{\rm H}$ is the Hermitian transpose;  $\left ( \cdot \right )^{\rm T}$ is the transpose.

\vspace{-4mm}
\section{System and Channel Models}
\begin{figure}[t]
\centering
\psfrag{A}[Bc][Bc][0.7]{$\mathrm{Base~ Station}$}
\psfrag{B}[Bc][Bc][0.7]{$\mathrm{Controller}$}
\psfrag{C}[Bc][Bc][0.7][-15]{$\mathrm{FRIS}$}
\psfrag{D}[Bc][Bc][0.7][-10]{$\mathrm{G}$}
\psfrag{E}[Bc][Bc][0.7][-15]{$\Psi$}
\psfrag{I}[Bc][Bc][0.7][-15]{$\mathrm{h}$}
\psfrag{F}[Bc][Bc][0.7]{$\mathrm{User}$}
\psfrag{G}[Bc][Bc][0.65]{$\mathrm{Preset~Position}$}
\psfrag{H}[Bc][Bc][0.65]{$\mathrm{Selected~Position}$}
\includegraphics[scale=0.24]{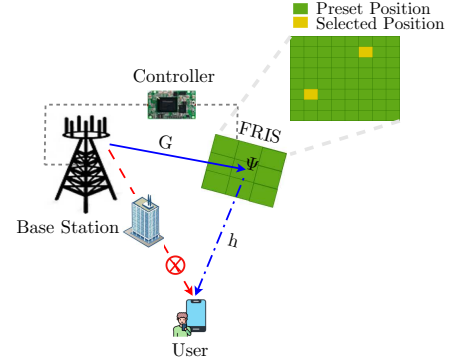}
\caption{System model for FRIS-enabled communication.
}
\label{SM}
\vspace{-4.5mm}
\end{figure}
We consider a \textcolor{black}{\ac{MISO}} communication system assisted by a \ac{FRIS}, as illustrated in Fig.~\ref{SM}, where the direct line-of-sight (LoS) path between the transmitter and the receiver is assumed to be blocked. In this setup, a base station (BS) equipped with an $L$-antenna uniform linear array communicates with a single-antenna user terminal through the aid of a FRIS composed of movable reflective elements. In the FRIS, the preset positions are uniformly distributed over a surface with total area $A_T=N\times A$, 
where $N = N_v \times N_h$ is the total number of preset positions available at the FRIS with $N_v$ and $N_h$ representing the number of \comment{predefined} locations per column and row, respectively. Moreover, $A=d_{\rm H}d_{\rm V}$ is the area of a preset FRIS element, where $d_{\rm H}$ is the horizontal width, $d_{\rm V}$ is the vertical height. Thus, the complete set of candidate (preset) locations across the FRIS surface is denoted by $\mathcal{C} = \{\mathbf{c}_1, \mathbf{c}_2, \dots, \mathbf{c}_N\}$, where each position is represented as a coordinate vector $\mathbf{c}_n = [c_{x,n}, c_{y,n}]^{\rm T}$, corresponding to the $n$-th location in the 2D deployment grid. This entire surface is divided into $M$ non-overlapping subareas $\{\mathcal{S}_1, \dots, \mathcal{S}_M\}$, with each subarea $\mathcal{S}_m$ for $m \in \{1,\dots,M\}$, exclusively assigned to one of the $M$ fluid elements that are allowed to be active\footnote{Each FRIS fluid element operates in one of two distinct modes: \textit{on} or \textit{off}. When in the \textit{off} mode, the element is connected to a matched load circuit and does not affect the incident electromagnetic signal. In contrast, in the \textit{on} mode, the element modifies the properties of the incoming wave—such as its phase—thereby contributing to signal shaping and performance optimization.}. Each fluid element is constrained to move only within its assigned subarea $\mathcal{S}_m$, where it can switch among a finite set of $N/M$ discrete preset positions arranged in a regular 2D grid within the subarea $\mathcal{S}_m$. The position of the $m$-th  fluid element is denoted by the vector $\mathbf{p}_m = [p_{x,m}, p_{y,m}]^{\rm T}$, and the set of all selected fluid elements is compactly expressed as $\mathcal{P} = [\mathbf{p}_1, \mathbf{p}_2, \dots, \mathbf{p}_M]$. In other words, $M$ fluid elements are selected from the total of $N$ preset positions available across the entire FRIS structure. Accordingly, the received signal at the user is a function of the selected positions \(\mathcal{P}\), and can be written as
\begin{equation}
\label{eq1}
y_{(\mathcal{P}) }= \sqrt{P} \, \mathbf{h}_{(\mathcal{P})}^{\mathrm{H}} \bm{\Psi}_{(\mathcal{P})} \mathbf{G}_{(\mathcal{P})} \mathbf{w} x + n,
\end{equation}
where $P$ is the transmit power at the BS, $x$ is the transmitted signal,
$n\sim \mathcal{C}\mathcal{N}(0,\sigma^2)$ is the additive white Gaussian noise with $\sigma^2$ power, and $\bf{w}$  denotes the BF vector at the BS. Also, ${\bf{G}}_{(\mathcal{P}) }= \left [ \bf{g}_1,\dots, \bf{g}_{L} \right ] \in \mathbb{C}^{M\times L}$ and ${\bf{h}_{(\mathcal{P})}}=\left [ h_{1},\dots,  h_{M} \right ]^{\rm H} \in \mathbb{C}^{M\times 1}$ denote the channel coefficients for the BS-to-FRIS and the FRIS-to-user paths, respectively.\footnote{Notice that, $\mathbf{h}$ and $\bf{w}$ are constructed based on the set $\mathcal{P}$ of selected fluid element positions.} Moreover, $\bm{\Psi}_{(\mathcal{P}) }=\diag\left ( e^{j\phi_{1}},\dots, e^{j\phi_{M}}\right )$ is the phase-shift matrix induced by the selected FRIS elements. For the sake of simplicity, we assume that the FRIS does not attenuate the reflected signals. 
The FRIS designs each reflecting element's phase-shifts so that the signals arriving at the user are aligned, i.e., $\phi_m=-\angle({ h}_{m}^{\rm H})-\angle({\bf{g}}_m{\bf{w}})$,
where ${ h}_{m}^{\rm H}$ is the $m$-th fluid element of ${\bf{h}}^{\rm H}$ and ${\bf{g}}_m$ is the $m$-th row vector of ${\bf{G}_{(\mathcal{P})}}$. Both $\mathbf{w}$ and $\bm{\phi}_m$ 
are subsequently optimized to maximize the end-to-end SNR. We assume that all links at the FRIS are under isotropic scattering (i.e., the Rayleigh model), as well as the indirect FRIS channels are spatially correlated. Thus, $ {\bf{h}}\sim \mathcal{C}\mathcal{N}\left ( \bm{0}_{ N},A \beta_{2} {\bf{R}_{(\mathcal{P})}} \vspace{0.8mm} \right )$, ${\bf{g}}_{\rm m}\sim \mathcal{C}\mathcal{N}\left ( \bm{0}_{M},A \beta_1 {\bf{R}_{(\mathcal{P})}} \right )$ for $m=\left \{ 1,\dots,M \right \}$,
where $\beta_1$, and $\beta_{2}$  encompass the average path loss attenuation for the links between BS-to-FRIS, and FRIS-to-user, respectively. Moreover, ${\bf{R}}_{(\mathcal{P})} \in \mathbb{C}^{M\times M}$ is the spatial correlation matrix on the FRIS. To calculate ${\bf{R}}_{(\mathcal{P})}$, we assume 
 \comment{Jakes'} model as a result of rich
scattering fading, so the $(a,b)$-th entry of the spatial correlation matrix ${\bf{R}}_{(\mathcal{P})}$ can be expressed as 
\vspace{-1mm} 
\begin{equation}\label{eq4}
{\left [{\bf{R}}_{(\mathcal{P})}  \right ] }_{a,b}= J_0 \left ( 2\left \| {\bf{u}}_a-{\bf{u}}_b \right \| /\lambda \right )     \hspace{2mm}  a,b=1,\dots,M,
\vspace*{-1mm} 
\end{equation}
where $\lambda$ denotes the carrier wavelength, ${\bf{u}}_\zeta =\left [ 0, {\rm{mod}}\left ( \zeta -1,{M_h}\right )d_{\rm H}, \left \lfloor \left (\zeta -1 \right )/{M_v} \right \rfloor d_{\rm V} \right ]^{T}$, $\zeta  \in \left \{ a,b \right \}$, and the terms $M_h$ and $M_v$ indicate the number of selected FRIS positions per row and column, respectively, of the entire metasurface, such that $M=M_h\times M_v$. Using the above formulations and \eqref{eq1}, the received \ac{SNR} at the user can be expressed as \vspace{-2mm}
\begin{align}\label{eqsnr}
\gamma_{(\mathcal{P}) }=&\overline{\gamma} \,| \mathbf{h}_{(\mathcal{P})}^{\mathrm{H}} \bm{\Psi}_{(\mathcal{P})} \mathbf{G}_{(\mathcal{P})} \mathbf{w}|^2,
\end{align}
where, we define $\overline{\gamma}=P/\sigma^2$ as the average transmit SNR for the user. 

\vspace{-1mm}
\section{Problem Formulation and Algorithm Design}
\comment{To maximize the SNR at the receiver side, the active (at the BS) and passive (PS at the FRIS) beamformers and the set of selected positions at the FRIS need to be jointly optimized.} The proposed design consists of two main components: $i)$ an EPSO-based algorithm to determine the optimal FRIS element positions, i.e., $\mathcal{P}$, and $ii)$ a joint PS and BF optimization based on a distributed approach for the selected elements.
\vspace{-1mm}
\subsection{Problem Formulation}
 We begin by optimizing the spatial configuration of the FRIS elements. The goal is to select the best \( M \) positions from the total of \( N \) available preset candidates across the FRIS surface in order to maximize the received SNR. Specifically, given the total set of $N$ candidate preset positions $\mathcal{C} = \{\mathbf{c}_1, \mathbf{c}_2, \dots, \mathbf{c}_N\}$  over the FRIS surface, we define a selected subset $\mathcal{P} \subset \mathcal{C}$ with $|\mathcal{P}| = M$ corresponding to the positions of the $M$ active fluid elements. Hence, the  optimization problem can be formulated as:
 \vspace{-1mm}
\begin{subequations}
\label{eq:opt_problem}
\begin{align}
(\rm{P1}): \quad & \max_{\mathcal{P} \subset \mathcal{C},\, |\mathcal{P}| = M} \quad \gamma_{(\mathcal{P})} \label{eq:opt_obj} \\
\text{s.t.} \quad
& \mathbf{p}_m \in \mathcal{S}_m, \quad \forall m = 1,\dots,M, \label{eq:pos_subarea} \\
& \|\mathbf{p}_m - \mathbf{p}_{m'}\|_2 \geq D, \quad \forall m \ne m', \label{eq:min_dist}
\end{align}
\end{subequations}
where $\mathbf{p}_m \in \mathbb{R}^2$ is the 2D coordinate of the $m$-th selected fluid element, the index ${m'}$ represents another selected fluid element that is distinct from $m$. 
Constraint~\eqref{eq:min_dist} enforces a minimum inter-element distance $D$ to avoid electromagnetic coupling or collision between elements.
\vspace{-1mm}
\subsection{Optimization Algorithm Design}
\subsubsection{Evolutionary Particle Swarm Optimization for Position Selection}

Due to the non-convex nature of the optimization problem formulated in (P1), conventional optimization techniques may struggle to find the global optimum. To overcome this challenge, we employ an E-PSO algorithm \cite{EPSO}, which enhances the global search capability of the traditional Particle Swarm Optimization (PSO) method by incorporating evolutionary operators such as mutation, elitism, and adaptive learning strategies. In E-PSO, each particle represents a candidate configuration $\mathcal{P}_i$ of $M$ fluid element positions selected from the set of $N$ available preset locations. The $i$-th particle is characterized by its current position vector $\mathbf{p}_i^{(t)}$ and velocity vector $\mathbf{v}_i^{(t)}$ at iteration $t$. The velocity and position updates are governed by the following \cite{EPSO}:
\vspace{-2mm}
\begin{align}
\mathbf{v}_i^{(t+1)} =\ & \delta \, \mathbf{v}_i^{(t)} + c_1 r_1 \left( \mathbf{p}_{i,\text{best}} - \mathbf{p}_i^{(t)} \right) + c_2 r_2 \left( \mathbf{p}_{\text{best}} - \mathbf{p}_i^{(t)} \right), \label{eq:pso_velocity} \\
\mathbf{p}_i^{(t+1)} =\ & \mathcal{E} \left( \mathcal{Z} \left( \mathbf{p}_i^{(t)} + \mathbf{v}_i^{(t+1)} \right) \right), \label{eq:pso_position}
\end{align}
where, $\delta$ is the inertia weight,
$c_1$ and $c_2$ are the cognitive and social acceleration coefficients, respectively, $r_1$ and $r_2$ are random scalars uniformly distributed in $[0,1]$, and $\mathcal{E}(\mathbf{z}) \triangleq \mathbf{z} + \boldsymbol{\xi}$, is the Gaussian mutation operator, with $\boldsymbol{\xi} \sim \mathcal{N}\!\left(\mathbf{0},\,\sigma_{\text{mut}}^{2}\mathbf{I}\right)$\footnote{Here, $\sigma_{\text{mut}}^{2}$ controls the mutation strength, and $\mathbf{I}$  $\in \mathbb{R}^{ 2\times 2}$ is the identity matrix ensuring  Gaussian perturbations in the 2D position space.
}. Also, $\mathbf{p}_{i,\text{best}}$ is the best position found so far by the $i$-th particle, $\mathbf{p}_{\text{best}}$ is the global best position among all particles, $\mathcal{Z}(\cdot)$ is a projection operator that ensures the updated positions are feasible and comply with the constraint defined in $(\rm{P1})$, including subarea assignments and minimum spacing between elements. Algorithm~\ref{alg:epso} outlines the proposed E-PSO-based method for selecting the optimal \comment{set of fluid} elements.
\begin{algorithm}[t]
\scriptsize
\caption{E-PSO-Based Position Optimization for FRIS}
\label{alg:epso}
\begin{algorithmic}[1]
\State \textbf{Input:} Total area $A_T = [X_{\min}, X_{\max}] \times [Y_{\min}, Y_{\max}]$, number of fluid elements $M$, minimum spacing $D$, power $P$, maximum iterations $T$, swarm size $S$, and E-PSO parameters $\delta$, $c_1$, $c_2$.
\State \textbf{Initialization:} Divide $A_T$ into $M$ subareas $\{\mathcal{S}_1, \dots, \mathcal{S}_M\}$ with spacing $\geq D$. Construct full candidate set $\mathcal{C} = \{\mathbf{c}_1, \dots, \mathbf{c}_N\}$. 
\State \quad Precompute: BS-to-FRIS channel $\mathbf{G} \in \mathbb{C}^{N \times L}$, FRIS-to-user channel $\mathbf{h}_u \in \mathbb{C}^{N \times 1}$, correlation matrix $\mathbf{R} \in \mathbb{C}^{N \times N}$.
\For {each fluid element $m = 1$ to $M$}
    \State Randomly initialize position $\mathbf{p}_m^{(0)}$ and velocity $\mathbf{v}_m^{(0)}$ within subarea $\mathcal{S}_m$.
   \For {each iteration $t = 1$ to $T$}
        \State Evaluate fitness with \eqref{eqsnr} for the current configuration $\mathcal{P}$.
        \State Update personal best $\mathbf{p}_{m,\text{best}}$ and global best $\mathbf{p}_{\text{best}}$.
        \State Update velocity using~\eqref{eq:pso_velocity}.
        \State Update position using~\eqref{eq:pso_position}.
  \EndFor
    \State Select $\mathbf{p}_m^{\star}$ as the best position found for element $m$.
    \If {$\|\mathbf{p}_m^{\star} - \mathbf{p}_{m'}^{\star}\| < D$ for any $m' < m$}
        \State Re-optimize $\mathbf{p}_m^{\star}$ within $\mathcal{S}_m$ to satisfy spacing.
    \EndIf
    \State Append $\mathbf{p}_m^{\star}$ to $\mathcal{P}^{\star}$.
\EndFor
\State \textbf{Output:} Optimized configuration $\mathcal{P}^{\star} = \{\mathbf{p}_1^{\star}, \dots, \mathbf{p}_M^{\star} \}$ and corresponding $\gamma_{(\mathcal{P}^{\star})}$ with final dimensions: $\mathbf{G} \in \mathbb{C}^{M \times L}$, $\mathbf{R} \in \mathbb{C}^{M \times M}$, $\mathbf{h} \in \mathbb{C}^{M \times 1}$.
\end{algorithmic}
\end{algorithm}
\subsubsection{Distributed Phase Shift and Beamforming Optimization}
Once the optimal set of positions $\mathcal{P}^\star$ has been determined, we jointly optimize the BS beamforming vector and the PS of the active fluid elements. The joint optimization problem can be formulated as:
\vspace{-1mm}
\begin{subequations}
\begin{align}
(\rm{P2}): \quad & \max_{\mathbf{w},\, \bm{\Psi}_{(\mathcal{P}^\star)}} \gamma(\mathcal{P}^\star) \label{eq:joint_opt_obj} \\
\text{s.t.} \quad & \|\mathbf{w}\|^2 \leq P_{\max}, \label{eq:joint_opt_c1} \\
& 0 \leq \phi_m < 2\pi,\quad \forall m = 1, \dots, M, \label{eq:joint_opt_c3},
\end{align}
\label{eq:joint_opt}
\end{subequations}
Here, $\gamma(\mathcal{P}^\star)$ denotes the end-to-end SNR under the fixed active element configuration $\mathcal{P}^\star$. Now, to solve $(\rm{P2})$, we adopt the low-complexity distributed algorithm based on alternating optimization \cite{MRT}. Specifically, the transmit BF at the BS and the phase shifts at the FRIS are optimized iteratively in an alternating manner, with one being fixed in each iteration, until convergence or a maximum number of iterations is reached. 
The distributed algorithm customized for the FRIS problem is described in Algorithm~\ref{alg:distributed}. Finally, after solving $(\rm{P1})$ and $(\rm{P2})$, 
the received signal at the user can be expressed as
\begin{equation}
y^\star_{(\mathcal{P^\star})} = \sqrt{P}\,\mathbf{h}_{(\mathcal{P}^\star)}^{\mathrm{H}} 
\bm{\Psi}_{(\mathcal{P}^\star)}^\star 
\mathbf{G}(\mathcal{P}^\star)\,\mathbf{w}^\star x + n,
\label{eqoptima}
\end{equation}
which represents the fully optimized end-to-end link. Consequently, the corresponding optimized SNR is given by
\begin{equation}
\label{SNRop}
\gamma^\star_{(\mathcal{P^\star}) } 
= \overline{\gamma}\,
\left| \mathbf{h}_{(\mathcal{P}^\star)}^{\mathrm{H}}
\bm{\Psi}_{(\mathcal{P}^\star)}^\star
\mathbf{G}(\mathcal{P}^\star)\,
\mathbf{w}^\star \right|^{2}=\overline{\gamma}\,z^2,
\end{equation}
where $z=\left|h^\star \right|$ denotes the optimized equivalent end-to-end channel.
\begin{algorithm}[t]
\scriptsize
\caption{Distributed Phase Shift and Beamforming Optimization}
\label{alg:distributed}
\begin{algorithmic}[1]
\State Initialize tolerance $\epsilon > 0$ and iteration index $k = 1$.
\State \textbf{Beamforming initialization:} Estimate $\mathbf{G}_{(\mathcal{P}^\star)} \in \mathbb{C}^{M \times L}$ and set
$\mathbf{w}^{(k)} = \mathbf{1}_L/\sqrt{L} \in \mathbb{R}^{L \times 1}$
\Repeat
    \State \textbf{Phase shift update:} With $\mathbf{w}^{(k)}$ fixed, compute
    \[
        \bm{\phi}_m^{(k+1)} = -\angle({ h}_{m}^{\rm H})-\angle({\bf{g}}_m{\bf{w}}^{(k)}),
    \]
    and set $\bm{\Psi}_{(\mathcal{P}^\star)}^{(k+1)}= \mathrm{diag}\left( e^{j\phi_1^{(k+1)}}, \dots, e^{j\phi_M^{(k+1)}} \right)$.
    \State \textbf{Beamforming update:} With $\bm{\Psi}_{(\mathcal{P}^\star)}^{(k+1)}$ fixed, calculate 
    \[  \mathbf{w}^{(k+1)} = \tfrac{\left ( \mathbf{h}_{(\mathcal{P}^\star)}^{\mathrm{H}} \bm{\Psi}_{(\mathcal{P}^\star)}^{(k+1)}\mathbf{G}_{(\mathcal{P}^\star)}  \right )^{\rm H}}{\norm{ \mathbf{h}_{(\mathcal{P}^\star)}^{\mathrm{H}} \bm{\Psi}_{(\mathcal{P}^\star)}^{(k+1)} \mathbf{G}_{(\mathcal{P}^\star)}  } } .
    \]
    \State $k \gets k + 1$.
\Until{fractional increase in $\gamma(\mathcal{P}^\star)$ $< \epsilon$ or maximum iterations reached}
\State \textbf{Output:} Optimized $\mathbf{w}^\star$ and $\bm{\Psi}^\star_{(\mathcal{P}^\star)}$.
\end{algorithmic}
\end{algorithm}
\vspace{-2mm}
\section{Performance Analysis}
\subsection{Outage Probability}
The \ac{OP} is considered to assess the performance of the FRIS,  which is defined as the probability that the received SNR  is less than a threshold rate $R$. Hence, the OP of the system can be formulated as
\begin{align}\label{eqOP}
 \text{OP}&=\Pr\left \{ \log_2\left ( 1+ \overline{\gamma}\,z^2\right )< R  \right \}   =F_{{z}}\left ( \sqrt{\tfrac{2^{R}-1}{\overline{\gamma}}} \right ) .
\end{align}
\vspace{-2mm}
\subsection{Statistical Approximation}
Here, the goal is to find an accurate yet tractable approximation of the effective channel magnitude $z$.  
To this end, we employ an unsupervised learning approach based on mixture clustering.  
In particular, we approximate the PDF of $z$ using a \textcolor{black}{finite} mixture of Nakagami-$m$ distributions, which provides a flexible and accurate representation even without labeled data or explicit knowledge of the propagation environment. Hence, let us consider a training set vector ${\bf{h}}^\star=\left \{ h^\star_i \right \}_{i=1}^{t_{sp}}$ consisting of $t_{sp}$ samples of $z$ given in \eqref{SNRop}.  
The distribution of $z$ can be approximated as
\vspace{-2mm}
\begin{align}\label{eq7}
f_{z}(r)= &\sum_{i=1}^{\mathcal{Q}}\alpha_{i}\varphi_{i}\left (r;m_i,\alpha_i\right )=\sum_{i=1}^{\mathcal{Q}}\alpha_{i}\tfrac{m_{i}^{m_{i}}r^{2m_i}e^{-\frac{m_{i}r^{2}}{\Omega_{i}}}}{2^{-1}\Gamma (m_{i})\Omega _{i}^{m_{i}} r} ,
\vspace{-3mm}
\end{align}
where $\alpha_{i}$ are the mixture weights satisfying $\sum_{i=1}^{\mathcal{Q}}\alpha_{i}=1$ with $0 \leq \alpha_{i} \leq 1$, $\mathcal{Q}$ denotes the total number of components in the mixture, $\Omega _{i}$ are the mean powers, $m_{i}$ are the fading parameters of the weighted PDFs.  
The parameters of the mixture model in \eqref{eq7} are estimated via the Expectation-Maximization (EM) algorithm \cite{JDVS}, which iteratively updates the membership coefficients and model parameters until convergence. Algorithm~\ref{algoritmo} summarizes the EM-based unsupervised clustering procedure for estimating the parameters $\alpha_i$, $\Omega_i$, and $m_i$.  
The mixture weights are initialized randomly in $[0,1]$, while the fading parameters are initialized via Maximum Likelihood Estimation (MLE).  
The stopping criterion is based on the relative tolerance between consecutive parameter updates, set experimentally to $1\times10^{-3}$.

\begin{algorithm}[t]
\scriptsize
\caption{EM-Based Unsupervised Clustering for End-to-End FRIS Channel}
\label{algoritmo}
\begin{algorithmic}[1]
\State \textbf{Input:} Training samples ${\bf h}^\star=\{h^\star_1,\ldots,h^\star_{t_{sp}}\}$, tolerance $\epsilon=10^{-3}$.
\State \textbf{Init:} Randomly initialize $\alpha_i^{(0)}\!\in[0,1]$ with $\sum_{i=1}^{2}\alpha_i^{(0)}=1$; initialize $m_i^{(0)}$, $\Omega_i^{(0)}$ via MLE; set $k\gets0$.
\While{relative changes of $m_i$ and $\Omega_i$ $>\epsilon$}
  \State \textbf{E–step:} For each $j=1,\ldots,t_{sp}$ and $i\in\{1,2\}$, compute memberships
  \[
  \tau_{ij}^{(k)}=\tfrac{\alpha_i^{(k)}\,\varphi_i\!\left(h^\star_j;m_i^{(k)},\Omega_i^{(k)}\right)}
  {\sum_{\ell=1}^{2}\alpha_\ell^{(k)}\,\varphi_\ell\!\left(h^\star_j;m_\ell^{(k)},\Omega_\ell^{(k)}\right)} .
  \]
  \State \textbf{M–step:} Update parameters using the memberships:
  \[
  \Omega_i^{(k+1)}=\tfrac{\sum_{j=1}^{t_{sp}}\tau_{ij}^{(k)} (h^\star_j)^2}{\sum_{j=1}^{t_{sp}}\tau_{ij}^{(k)}}, \quad
  \alpha_i^{(k+1)}=\tfrac{\sum_{j=1}^{t_{sp}}\tau_{ij}^{(k)}}{t_{sp}},
  \]
  \[
  m_i^{(k+1)}=\tfrac{1+\sqrt{1+\frac{4\Delta_i^{(k)}}{3}}}{4\,\Delta_i^{(k)}}, \quad
  \Delta_i^{(k)}=\tfrac{\sum_{j=1}^{t}\tau_{ij}^{(k)}\!\left[\log\!\big(\Omega_i^{(k)}\big)-\log\!\big((h^\star_j)^2\big)\right]}
  {\sum_{j=1}^{t}\tau_{ij}^{(k)}}.
  \]
  \State $k\gets k+1$.
\EndWhile
\State \textbf{Output:} Mixture parameters $\{\alpha_i,m_i,\Omega_i\}_{i=1}^\mathcal{Q}$ and fitted PDF $f_{z}(r)=\sum_{i=1}^{\mathcal{Q}}\alpha_i\,\varphi_i(r;m_i,\Omega_i)$.
\end{algorithmic}
\end{algorithm}

\begin{proposition}\label{Propo1}
The approximate OP expression of the FRIS system is given by
\end{proposition}
\vspace{-9mm}
\begin{align}\label{eq11}
   \text{OP}= 1-\sum_{i=1}^{\mathcal{Q}}\frac{\Gamma\left (m_i,\frac{m_i\left ( 2^{R_{\rm th}}-1 \right )}{\Omega_i P_{\rm T}/\sigma_{\widetilde{w}}^2}  \right )}{\Gamma\left ( m_i \right )},
\end{align}
where the shape parameters are estimated with the help of Algorithm~\ref{algoritmo}.
\begin{proof}
    From \eqref{eqOP}, the OP in \eqref{eq11} can be obtained directly from the CDF of \eqref{eq7}.
\end{proof}
\section{Numerical results and discussions} \label{sect:numericals}

In this section, we present numerical results to validate the proposed analytical framework and to quantify the performance gains of the FRIS over RIS architectures. The analysis leverages the previously derived channel modeling and optimization procedures. Specifically, two FRIS configurations are considered: $i)$ FRIS with only Selected Positions Optimized (SPO), and $ii)$ FRIS where the element positions, transmit BF, and PS are all jointly optimized (SPO+BF+PS). The FRIS is benchmarked against two RIS layouts. 
$a)$ A \emph{Conventional RIS} layout, where $M$ fixed RIS elements are located at the centers of the subregions corresponding to the FRIS preset grid, spanning the full physical aperture area. 
$b)$ A \emph{Compact RIS} layout, where $M$ elements are arranged in a contiguous rectangular array with uniform inter-element spacing $\leq \lambda/2$. The resulting physical aperture is therefore significantly smaller than that of the full FRIS configuration.
For a fair comparison, both RIS configurations are implemented with the same total number of elements $M$ as in the FRIS scenario, and can operate either with random PS or with both BF and PS jointly optimized. Unless otherwise stated, the FRIS parameters are set as follows: carrier frequency $f_c = 2.4$~GHz (yielding $\lambda=0.125$~m), correlation matrix constructed with $d_{\rm H} = d_{\rm V} = \lambda/3$, aperture size $A_T = 2\,\text{m} \times 2\,\text{m}$ resulting in $N_{\rm v} = N_{\rm h} = 48$, path-loss exponents $\beta_1=\beta_2=-40$~dB, and SNR threshold $R = 1$~bps/Hz for all OP evaluations. The E-PSO parameters are set to $S=150$, $T=20$, $\delta=0.6$, $c_1=1.8$, and $c_2=1.8$. 
For the EM algorithm, $t_{\rm sp} = 10^4$ realizations are generated for the training set in all instances, with $\mathcal{Q} = 2$ as the number of components in the mixture. Monte Carlo simulations over the true channel realizations are carried out to validate the accuracy of the proposed analytical framework. For the sake of analytical comparison, the Moment Matching (MoM) and
Kolmogorov–Smirnov (KS) approaches are incorporated as benchmark methods in the OP analysis.
\begin{figure*}[ht!]
\vspace{-5.5mm}
\centering
\psfrag{A}[Bc][Bc][0.5]{$\rho=5$ dB}
\psfrag{B}[Bc][Bc][0.5]{$\rho=-5$ dB}
\psfrag{C}[Bc][Bc][0.5]{no EMI}
\psfrag{D}[Bc][Bc][0.5]{$\rho=5$ dB}
\subfigure[FRIS spatial configuration with $M = 16$ and $L=2$.]{\includegraphics[width=0.32\textwidth, height=0.18\textheight]{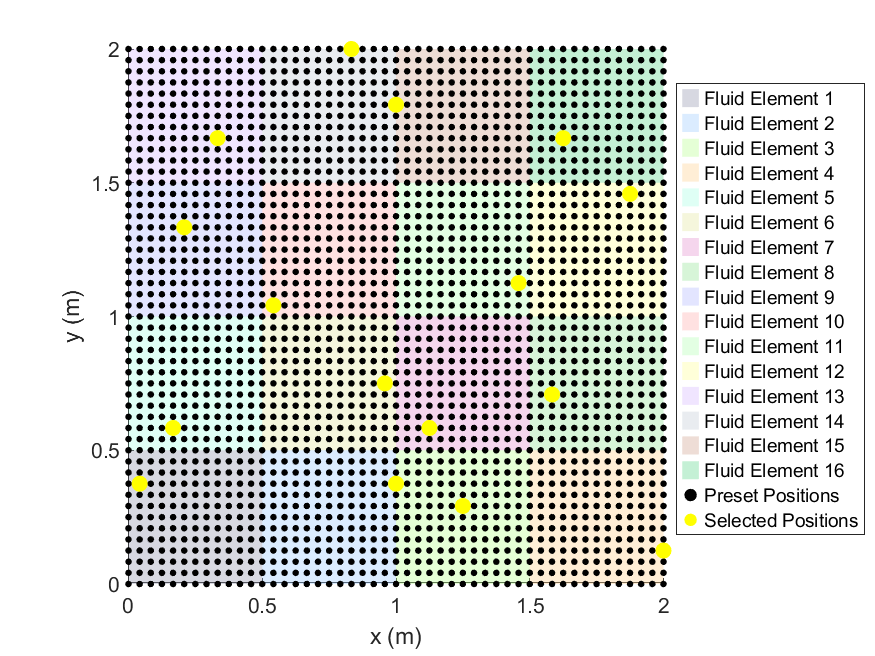}} 
\psfrag{E}[Bc][Bc][0.5]{$\rho=20$ dB}
\psfrag{A}[Bc][Bc][0.5]{$\rho=20$ dB}
\psfrag{B}[Bc][Bc][0.5]{$\mathrm{\textit{N}=324, 144}$}
\psfrag{C}[Bc][Bc][0.5]{$\sigma_{\text{EMI},\mathrm{B}}^2=-45$ dBm}
\psfrag{D}[Bc][Bc][0.5]{$\mathrm{\textit{N}=144, 324}$}
\subfigure[OP vs. $\overline{\gamma}$ for $M=16$ fluid elements and $L \in \{1,5\}$. ]{\includegraphics[width=0.32\textwidth]{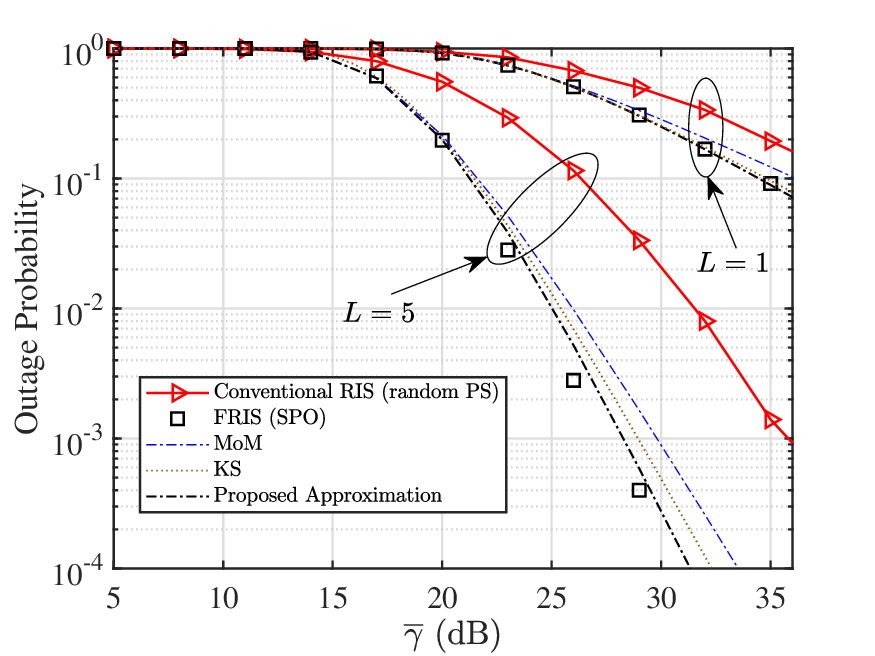}}
\psfrag{A}[Bc][Bc][0.5]{$\mathrm{\lambda/2}$}
\psfrag{B}[Bc][Bc][0.5]{$\mathrm{\lambda/5}$}
\psfrag{C}[Bc][Bc][0.5]{$\mathrm{\lambda/3}$}
\subfigure[OP vs. $\overline{\gamma}$ for $M=16$ and $L=3$.]{\includegraphics[width=0.32\textwidth]{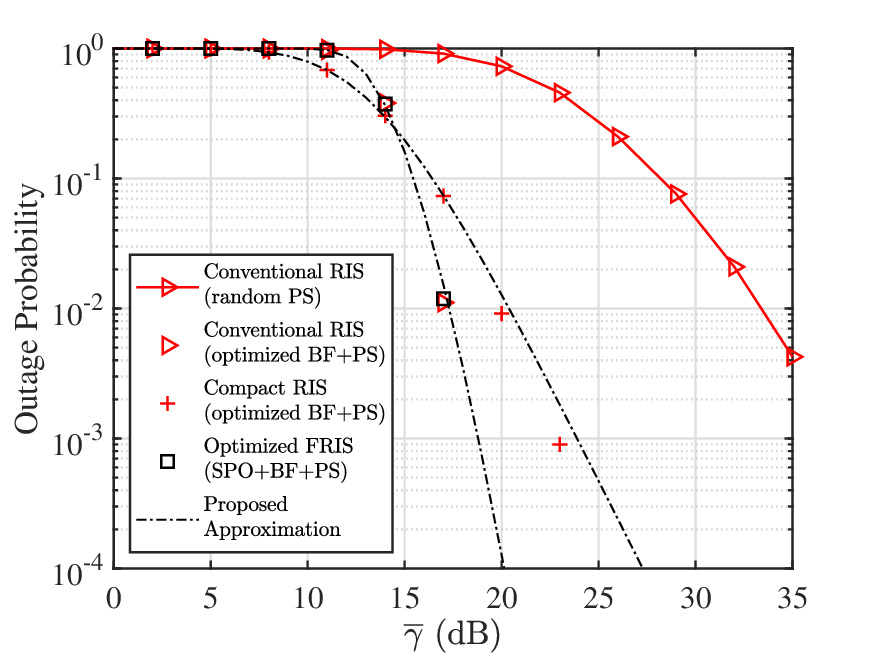}}
    \caption{Spatial configuration and OP achieved for FRIS-aided Multiple-Input Single-Output (MISO). In all figures, the dashed lines represent analytical solutions, while markers correspond to MC simulations. Also, for a fair comparison, the RIS is configured with the same number of elements $M$ as the FRIS in all cases. }
    \label{figen}
    \vspace{-4mm}
\end{figure*}
\begin{figure}[t]
\vspace{-2.5mm}
\centering
\psfrag{A}[Bc][Bc][0.5]{$\rho= 30, 25, 20$ dB}
\psfrag{B}[Bc][Bc][0.5]{$\rho=40$ dB}
 \includegraphics[width=60mm]{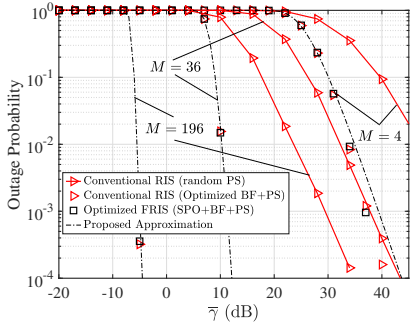}
\caption{OP vs. $\overline{\gamma}$ for $L=2$, and $M \in \{4,36,196\}$ fluid elements. Markers denote MC simulations whereas the solid lines represent analytical solutions.}
\label{fig5f}
\vspace{-4.5mm}
\end{figure}

Figure~\ref{figen}a illustrates the spatial configuration of the FRIS for the case of $M=16$ fluid elements and $L=2$ transmit antennas. The $N$ candidate elements are uniformly distributed across the $M$ subareas, ensuring an even spatial coverage of the deployment region. Each subarea corresponds to a possible location for one active fluid element, and the optimal choice is determined through the proposed E-PSO framework for solving Problem~P1. In the figure, the yellow markers denote the positions selected by E-PSO, representing the geometric configuration of the FRIS after the position optimization stage. 

In Fig.~\ref{figen}b, we compare the FRIS system with \textit{only} position optimization, i.e., SPO, against a conventional RIS with \textit{random} PS for $M=16$ fluid elements and $L \in \{1,5\}$. Here, we focus on  neither FRIS nor RIS employ phase shift nor beamforming optimization, thus isolating the effect of the geometric configuration. From curves, it is \comment{observed} that employing FRIS with the optimized geometry obtained from solving\footnote{\textcolor{black}{Effectiveness of E-PSO has been verified offline against a brute-force approach.}} $\rm{P1}$ provides a clear performance gain over a conventional RIS without phase design. This gain, however, will be further contrasted against an optimized RIS as we will see later. In addition, the curves also includes the proposed EM-based approximation, as well as the MoM and KS analytical benchmarks. Here, it is observed that the proposed EM approach exhibits the best fitting to the \comment{true} OP metric.

In Fig.~\ref{figen}c, we investigate the performance of the FRIS with full optimization (SPO+BF+PS) against an optimized RIS design (BF+PS). The configuration parameters are set to $M = 16$ fluid elements and $L = 3$ BS antennas. Here, the primary goal is to assess whether the spatial configuration gain of the FRIS, observed in Fig.~\ref{figen}b, can provide an advantage over an \textit{optimized} RIS with a carefully designed PS and BF strategy. For reference, a Conventional RIS (random PS) is also included in the analysis. From all plotted curves, it can be observed that the OP performance of the optimized FRIS (SPO+BF+PS) and the optimized Conventional RIS (BF+PS) is indistinguishable across the entire average transmit SNR range. This finding suggests that, when an RIS benefits from a well-designed BF+PS configuration, the additional spatial optimization provided by the FRIS does not translate into meaningful OP gains. In other words, the deployment of FRIS in such conditions is not strongly justified from a performance standpoint.  On the other hand, as expected, the Conventional RIS exhibits the poorest OP performance, which is attributed to the absence of any PS or BF optimization. Furthermore, the results reinforce the notion that the main advantage of FRIS lies in scenarios where PS optimization is limited, in which case the spatial configuration becomes a decisive factor in enhancing system performance. 
It is also worth noting that the FRIS outperforms the Compact RIS (optimized BF+PS) across the SNR range. This improvement stems from the fact that, unlike the Conventional RIS, the elements in the Compact RIS are placed much closer together, which increases the spatial correlation and reduces channel diversity. The FRIS, with its optimized spatial configuration, mitigates these correlation effects, thereby achieving superior outage probability performance compared to the Compact RIS.

In Figure~\ref{fig5f}, we show the OP versus the average SNR $\overline{\gamma}$ for both the FRIS and RIS under full optimization, considering different numbers of elements $M \in \{4, 36, 196\}$ and $L=3$.
Here, we investigate whether the number of elements influences the relative OP performance gain of FRIS over RIS. From the curves, it is observed that the optimized FRIS (SPO+BF+PS) and the optimized conventional RIS (BF+PS) exhibit nearly identical OP performance across all tested values of $M$. This indicates that, when both systems employ optimal PS and transmit BF, the spatial configuration advantage of FRIS does not \textcolor{black}{improve performance} over RIS, regardless of the number of fluid elements considered. 

\vspace{-1mm}
\section{Conclusions}
This paper presented a comparative study of FRIS and RIS systems under different optimization strategies, including position selection, BF, and PS design. Results show that FRIS with SPO significantly outperforms conventional RIS without optimization, confirming the benefit of spatial reconfiguration \comment{in the absence of PS optimization}. However, when both systems employ optimal BF and PS, their OP performance becomes nearly identical, indicating that the spatial advantage of FRIS is not relevant  \comment{over fully optimized conventional RIS}. 

\bibliographystyle{ieeetr}
\bibliography{bibfile}

@ARTICLE{FASRIS,
  author={Rostami Ghadi, Farshad and others},
  journal={IEEE Wireless Commun. Lett.}, 
  title={{On Performance of RIS-Aided Fluid Antenna Systems}}, 
  year={2024},
  volume={13},
  number={8},
  pages={2175-2179},
  doi={10.1109/LWC.2024.3405636}}

@article{Farshad,
  title={Performance Analysis of Wireless Communication Systems Assisted by Fluid Reconfigurable Intelligent Surfaces},
  author={Farshad Rostami Ghadi and others},
  journal={IEEE Wireless Comun. Lett.}, 
  year={2025},
  volume={},
  number={},
  pages={1-1},
}

@article{Xiao,
  author={Xiao, Han and others},
  journal={IEEE Wireless Commun. Lett.}, 
  title={{Fluid Reconfigurable Intelligent Surfaces:Joint On-Off Selection and Beamforming With Discrete Phase Shifts}}, 
  year={2025},
  volume={},
  number={},
  pages={1-1},
  doi={10.1109/LWC.2025.3587070}
}

@incollection{EPSO,
  title={{EPSO: Evolutionary Particle Swarms}},
  author={Miranda, V. and Keko, H. and Jaramillo, A.},
  booktitle={Advances in Evolutionary Computing for System Design},
  series={Studies in Computational Intelligence},
  volume={66},
  pages={139--167},
  year={2007},
  publisher={Springer, Berlin, Heidelberg},
  doi={10.1007/978-3-540-72377-6_6}
}

@article{Rostami,
  title={{FIRES: Fluid Integrated Reflecting and Emitting Surfaces}},
  author={Ghadi, Farshad Rostami and others},
  journal={IEEE Wireless Comun. Lett.}, 
  year={2025},
  volume={},
  number={},
  pages={1-1},
  doi={10.1109/LWC.2025.3602219}}

@article{Abdelhamid,
  title={A First Look at the Performance Enhancement Potential of Fluid Reconfigurable Intelligent Surface},
  author={Abdelhamid Salem and others},
  journal={arXiv preprint arXiv:2502.17116},
  year={2025},
  url={https://arxiv.org/pdf/2502.17116}
}

@ARTICLE{Ref2,
  author={Wong, Kai-Kit and Shojaeifard, Arman and Tong, Kin-Fai and Zhang, Yangyang},
  journal={IEEE Trans. Wireless Commun.}, 
  title={Fluid Antenna Systems}, 
  year={2021},
  volume={20},
  number={3},
  pages={1950-1962},
  doi={10.1109/TWC.2020.3037595}}

@ARTICLE{JDVS,
  author={Sánchez, José David Vega and Urquiza-Aguiar, Luis and Paredes, Martha Cecilia Paredes and López-Martínez, F. Javier},
  journal={IEEE Wireless Commun. Lett.}, 
  title={{Expectation-Maximization Learning for Wireless Channel Modeling of Reconfigurable Intelligent Surfaces}}, 
  year={2021},
  volume={10},
  number={9},
  pages={2051-2055},
  doi={10.1109/LWC.2021.3091840}}

@INPROCEEDINGS{MRT,
  author={Wu, Qingqing and Zhang, Rui},
  booktitle={2018 IEEE Global Communications Conference (GLOBECOM)}, 
  title={Intelligent Reflecting Surface Enhanced Wireless Network: Joint Active and Passive Beamforming Design}, 
  year={2018},
  volume={},
  number={},
  pages={1-6},
  doi={10.1109/GLOCOM.2018.8647620}}

\end{document}